\documentclass[12pt,superscriptaddress,a4paper,showpacs,showkeys,pra]{revtex4}
\usepackage{graphicx,subfigure}

\begin{document}

\title{Mixing and Decoherence to Nearest Separable States}

\author{Avijit Lahiri}
\email{lahiri@cal2.vsnl.net.in}
\affiliation{Vidyasagar Evening College, Kolkata 700 006, India.}
\author{Gautam Ghosh}
\email{gautam.ghosh@saha.ac.in}
\affiliation{Saha Institute of Nuclear Physics, 1-AF Bidhannagar, Kolkata 700 064, India.}
\author{Sankhasubhra Nag}
\email{sankha@theory.saha.ernet.in}
\affiliation{Sarojini Naidu College for Women, Kolkata 700 028, India.}
\begin{abstract}
We consider a class of entangled states of a quantum system (S) and a second system (A) where pure states of the former are correlated with mixed states of the latter, and work out the entanglement measure with reference to the nearest separable state. Such `pure-mixed' entanglement is expected when the system S interacts with a macroscopic measuring apparatus in a quantum measurement, where the quantum correlation is destroyed in the process of environment-induced decoherence whereafter only the classical correlation between S and A remains, the latter being large compared to the former. We present numerical evidence that the entangled S-A state drifts towards the nearest separable state through decoherence, with an additional tendency of equimixing among relevant groups of apparatus states. 
\end{abstract}
\maketitle

\section{Introduction}
\label{Introduction}
A measurement involves an environment (E) in addition to a measured system (S) and a measuring apparatus (A). The apparatus has to be considered classical in the sense that it is an open system with a large number of degrees of freedom interacting weakly with an enormous number of environmental variables \cite{Venugopalan1999}. This interaction destroys phase correlations between apparatus states as also those  between the system and the apparatus resulting from the pre-measurement interaction \cite{Zurek2003}. The latter correlates specific system states with specific groups of apparatus states termed pointer states which are, in reality, macroscopic states, being mixtures of large numbers of apparatus microstates \cite{Wigner1963,Peres1993}. Environment-induced decoherence destroys the phase or quantum component of this correlation wherein only the classical component survives, embodying the correct measurement statistics.
\vskip .5cm
\noindent
The fact that the apparatus is a macroscopic system requires one to consider mixed states of the apparatus rather than pure ones and, following a recent work \cite{Lahiri2007}, we examine the question of a general characterization of the environment-induced decoherence in the measurement process  taking into account the involvement of mixed apparatus states \cite{Vedral2003}. 
\vskip .5cm
\noindent
More precisely, we refer to a correlation between pure system states and mixed apparatus states, designated below as `pure-mixed' entanglement in contrast to the `pure-pure' entanglement considered more commonly, the former having a structure similar to the Schmidt-decomposed representation of the latter with the important difference, however, that the arbitrariness in the Schmidt decomposition \cite{Zurek2003,Peres1993} for degenerate Schmidt coefficients is removed in the pure-mixed case.  This, though, does not completely eliminate the so-called preferred basis problem since the question remains as to the principle governing the course of the environment-induced decoherence process whereby one particular disentangled system-apparatus state, namely the one embodying the correct measurement statistic, results from among the set of all possible separable states. The characterization of the decoherence process referred to above appears to provide us with such a principle, at least in a limited context.
\vskip .5cm
\noindent
We consider below finite dimensional state spaces for the systems S, A, and E for the sake of convenience, with quite arbitrary interactions between the apparatus and the environment, represented by real symmetric matrices with randomly selected elements, and present evidence that the principle is a simple and general one: the entangled system-apparatus state evolves by decoherence to the {\it nearest} separable state, wherein the quantum part of the information is erased, leaving intact the classical part \cite{Lahiri2007}.
\vskip .5cm
\noindent
The fact that the apparatus-environment interaction is chosen to be a randomly selected one implies the absence of what is known as einselection \cite{Zurek1981,Zurek1982}, and hence raises the question of stability of the pointer states against environmental perturbations. We assume below that the pointer states of the apparatus have well-defined energy values so as to possess such stability - a feature that is possibly linked with the macroscopic nature of the apparatus \cite{Peres1993,Frasca2006}.

\vskip .5cm
\noindent
This article is organized as follows. In section \ref{pme} below we consider a system S with a 2D state space and propose a density operator $\rho$ for the system-apparatus entangled state of the pure-mixed type, arrived at through a purification procedure involving a notional auxiliary system. Assuming that a pre-measurement interaction \cite{Zurek2003} results in such a state, we characterise $\rho$ in terms of its eigenvalues and those of its partial transpose \cite{Peres1993}, noting that precisely one among the latter set of eigenvalues is negative. This single negative eigenvalue (generalization for a system with a state space of a higher dimension will be stated) will later be found to be a good index to monitor the process of decoherence wherein it will be seen to remain negative throughout the process, reaching the value zero at the end of it.
\vskip .5cm
\noindent
In section \ref{nss}, we compare the above pure-mixed entangled state with one of the pure-pure type for which a Schmidt decomposition is possible, and point out that the arbitrariness inherent in the latter for the {\it degenerate} pure-pure case is generically absent in the former owing to the different numbers of independent microstates in the various groups of apparatus macrostates (see below), indicating its implication in the context of the so-called preferred-basis problem in quantum measurements.
\vskip .5cm
\noindent
We demonstrate that the density operator  $\rho^*$ obtained from $\rho$ by the deletion of off diagonal blocks is the nearest disentangled state (in the sense of \cite{Vedral1998}) to $\rho$ when an appropriate measure of distance is adopted. We also show through a solvable example how the degree of entanglement is reduced drastically when mixed  states of the apparatus involving large numbers of microstates are considered. This provides the basis for the principle, referred to above, characterising the decoherence process outlined in \cite{Lahiri2007} and described in the next section.
\vskip .5cm
\noindent
Section \ref{dcp} provides the background needed for the numerical computation of the time evolution. Results of the computation are presented in section \ref{nume} to show that the decoherence process involves two quite general trends: (a) mixing between each bunch of microstates making up a pointer state of the apparatus so that a microcanonical distribution among these states is brought about, and (b) evolution towards the nearest separable system-apparatus state. While there occurs Brownian-like fluctations in the density matrix during the course of this evolution it still succeeds in `seeking out' this separable state owing to the fact that only a small amount of information is dissipated into the environment before the classically correlated state of the composite system is arrived at. We present illustrative numerical evidence of the decoherence process wherein an initial state with entanglement of the pure-mixed type ends up as a disentangled state with the classical S-A correlation remaining intact. Section \ref{ocr} is devoted to a summary appraisal of our results.

\section{Pure-mixed entanglement}
\label{pme}
\noindent
We begin with a density operator, assumed to be the end product of the pre-measurement process (see, however, section \ref{ocr}), in which the orthogonal states $|s_1\rangle$, $|s_2\rangle$ of a two-state system (S), being eigenstates of an observable, say, $\hat{S}\equiv s_1|s_1\rangle\langle s_1|+s_2|s_2\rangle\langle s_2|$, get correlated with two mixed apparatus states, say, $\rho_a^{(A)}$  and $\rho_b^{(A)}$ respectively, each corresponding to a definite value of the relevant pointer variable:

\begin{eqnarray}
\rho=|c_1|^2|s_1\rangle \langle s_1|\otimes \rho_a^{(A)}+
|c_2|^2|s_2\rangle \langle s_2|\otimes \rho_b^{(A)}
+c_1c_2^*|s_1\rangle \langle s_2|\otimes
 |\phi_a^{(A)}\rangle\langle\phi_b^{(A)}|\nonumber\\+c_1^*c_2|s_2\rangle \langle s_1|\otimes|\phi_b^{(A)}\rangle\langle\phi_a^{(A)}|.
\end{eqnarray}
\noindent
Here $c_i~(i=1,2)$ are the amplitudes of the states states $|s_i\rangle$ of the system under measurement, and the relevant mixed states of the apparatus are, say,
\begin{eqnarray}
\rho_a^{(A)}={\sum_{i=1}^{N_1}p_i|a_i\rangle \langle a_i|},
\end{eqnarray}
\begin{eqnarray}
\rho_b^{(A)}={\sum_{i=1}^{N_2}q_i|b_i\rangle \langle b_i|},
\end{eqnarray}
\noindent
made up of orthonormal microstates $|a_i\rangle$, $|b_j\rangle~~(i=1,\ldots,N_1,~j=1,\ldots,N_2)$ belonging to two subspaces of dimensions $N_1$ and $N_2$ for the apparatus. Normalization is ensured by requiring that the sets of weights $|c_i|^2~(i=1,2)$, $p_i~(i=1,\ldots,,N_1)$, and $q_j~(j=1,\ldots,N_2)$ are each separately normalized. Thus, $|c_1|^2$ and $|c_2|^2$ provide the measurement statistics in the projective measurement of the observable $\hat{S}$ of S. 
\vskip .5cm
\noindent
The off-diagonal terms in (1) make the density operator entangled, where the vectors $|\phi_a^{(A)}\rangle$ and $|\phi_b^{(A)}\rangle$ in these off-diagonal terms are given by,

\begin{eqnarray}
|\phi_a^{(A)}\rangle={\sum_{i=1}^{N_1}p_i|a_i\rangle},
\end{eqnarray}
\noindent and
\begin{eqnarray}
|\phi_b^{(A)}\rangle={\sum_{i=1}^{N_2}q_i|b_i\rangle}.
\end{eqnarray}
\vskip .5cm
\noindent
To show that the $\rho$ given by (1) is a legitimate density operator (i.e., is a positive operator with unit trace) we refer to the following purification procedure. We introduce an auxiliary system (say, H) with orthonormal basis states $|e_{lk}\rangle$ ($l=1...N_1, k=1...N_2$) and write down a pure state for the composite system made up of S, A, and H, with a Schmidt decomposition 
\begin{eqnarray}
|\psi\rangle=c_1|s_1\rangle\otimes {\sum_{i=1}^{N_1}\sqrt {p_i} |a_i\rangle}\otimes |\alpha_i\rangle +c_2|s_2\rangle\otimes {\sum_{i=1}^{N_2}\sqrt{q_i} |b_i\rangle}\otimes |\beta_i\rangle,
\end{eqnarray}
\noindent where the components of $|\alpha_i\rangle$ ($i=1\ldots,N_1$) and $|\beta_j\rangle$ ($j=1\ldots,N_2$) are given by,
\begin{eqnarray}
\langle e_{lk}|\alpha_i\rangle=\sqrt{Q_k}\delta_{il},
\end{eqnarray}
\noindent
and
\begin{eqnarray}
\langle e_{lk}|\beta_j\rangle=\sqrt{P_l}\delta_{jk}.
\end{eqnarray}
\noindent
This implies that
both $|\alpha_i\rangle$ and $|\beta_j\rangle$ form orthonormal sets provided that ${\sum_{k=1}^{N_2}}Q_k={\sum_{l=1}^{N_1}}P_l=1$, i.e., the $P$'s and $Q$'s are any two appropriately chosen sets of weights. We can thus choose, as a special case, the $P$'s and $Q$'s as the $p$'s and $q$'s respectively.
\vskip .5cm
\noindent
The $\rho$ given by (1) can now be seen to be the reduced density operator resulting from the pure state density operator $|\psi\rangle \langle\psi|$, by tracing over the auxiliary system states. Generalization to a system (S) with a state space of more than two dimensions and an apparatus with a correspondingly larger number of groups of microstates is straightforward.
\vskip .5cm
\noindent
One can have further information on $\rho$ from its structure (referred to the product basis formed with vectors $|s_1\rangle$, $|s_2\rangle$ for S, and $|a_i\rangle$, $| b_j\rangle$ for A) which shows that the rows and columns numbered $N_1+1$ to $2N_1+N_2$ are identically zero, providing us with $(N_1+N_2)$ trivially zero eigenvalues.  If we delete these rows and columns, the resulting collapsed matrix ($\rho_c$) has a particularly simple form, viz., it has two diagonal blocks which are diagonal matrices with elements $|c_1|^2p_i$ and $|c_2|^2q_j$ respectively and transposed conjugate off diagonal blocks with elements of the form $c_1c_2^*p_i q_j$, i.e., it takes the form,
\vskip .5cm
\begin{equation}
	\rho_c=
\left(\begin{array}{cccc}
	|c_1|^2[p_i\delta_{ij}] & c_1c_2^*[p_iq_j]\\ c_1^*c_2[q_jp_i] & |c_2|^2[q_j\delta_{ij}] 
\end{array}\right)
\end{equation}
where the entries are understood to be submatrices with appropriate dimensions.
\noindent  We can now calculate the determinant which turns out to be,
\begin{equation}
\mathrm{det}(\rho_c)=(|c_1|)^{2N_1}(|c_2|)^{2N_2}\prod_l p_l\prod_k q_k (1-\sum_i p_i\sum_j  q_j),
\end{equation}
\noindent where the appropriate limits for the sums and products are understood. Since the $p$'s and $q$'s separately sum to unity, the determinant is zero, and thus there arise zero eigenvalues of $\rho_c$ whose number can now be determined from a consideration of the linear term in the characteristic polynomial whose roots are the eigenvalues. It is obvious that the coefficient of $\lambda$ in $\prod_j(\lambda-\lambda_j)$ is zero only when at least two of the eigenvalues are zero. However, this coefficient is just the negative of the sum of the cofactors of the diagonal elements of $\rho_c$. Since the  deletion of the row and column corresponding to a diagonal element of $\rho_c$ is equivalent to eliminating one of the $p$'s or $q$'s which now no longer add to unity in (10), it is apparent that the said cofactors are all positive, and there is only one zero eigenvalue (the eigenvector corresponding to this zero eigenvalue is a column whose first $N_1$ entries are $c_1c_2^*$ while the rest are $-|c_1|^2$). The total number of non-zero eigenvalues of $\rho$ is thus seen to be $(N_1+N_2-1)$.   

\vskip .5cm
\noindent
The partial transpose of $\rho$ with respect to, say S, is defined by,
\begin{equation}
\langle s_i,\alpha|\rho^{PT}|s_j,\beta\rangle=\langle s_j,\alpha |\rho|s_i,\beta\rangle \hspace{1cm}   (i,j=1,2)
\end{equation}
\noindent and is of interest because of the result \cite{Peres1993} that for bipartite systems the necessary condition for a density matrix to be separable is that it has positive partial transpose (PPT); for entangled states however the partial transpose may or may not be positive. It is also known that the condition is sufficient for $2\otimes 2$ and $2\otimes 3$ dimensional composite systems \cite{Horodecki1996}. In our case, $\rho^{PT}$ is a block diagonal matrix with two of the blocks already diagonal, providing us with the eigenvalues $|c_1|^2p_i$ ($i=1\ldots,N_1$) and $|c_2|^2q_j$ ($j=1\ldots,N_2$). The other block which we call the central block, is a matrix which has identically zero diagonal blocks, and transposed conjugate off diagonal blocks with elements of the form $c_1c_2^*p_i q_j$. In other words,
\vskip .5cm
\noindent
\begin{equation}
	\rho^{PT}=
\left(\begin{array}{cccc}
	|c_1|^2[p_i\delta_{ij}] & 0 & 0 & 0\\0 & 0 & c_1^*c_2[q_jp_i] & 0\\0 &  c_1c_2^*[p_iq_j] & 0 &0\\ 0 & 0 & 0 &|c_2|^2[q_i\delta_{ij}]
\end{array}\right).
\end{equation}

\vskip .5cm
\noindent
The square of the central block is therefore block diagonal with  diagonal blocks of the form $MM^{\dagger}$ and $M^{\dagger}M$ where $M$ stands for the matrix $c_1^*c_2[q_j p_i]$. Defining unit vectors $|\psi\rangle$ having components $q_j/(\sum_k q_k^2)$ and $|\phi\rangle$ having components $p_i/(\sum_l p_l^2)$ we can write,
\begin{equation}
MM^{\dagger}=|c_1|^2|c_2|^2(\sum_i p_i^2 \sum_j q_j^2)|\psi\rangle \langle\psi|
\end{equation}
and
\begin{equation}
M^{\dagger}M=|c_1|^2|c_2|^2(\sum_i p_i^2 \sum_j q_j^2)|\phi\rangle \langle\phi|.
\end{equation}
The only nonzero eigenvalue of the projection operators is one and all other eigenvalues are zero. Thus the central block has only two nonzero eigenvalues viz. $\pm |c_1||c_2|\sqrt {\sum{p_i^2}\sum{q_j^2}}$, which means that there are now a total of $(N_1+N_2+2)$ non-zero eigenvalues out of which precisely one is negative. Our numerical investigation (section \ref{nume}) will show how this negative eigenvalue approaches zero as the decoherence process develops. Generalization to the situation where the state space of S has a dimension larger than two, with a correspondingly larger number of groups of microstates of A, is straightforward. For instance, with a 3D state space of S, and with weights $p_i$, $q_j$, $r_k$ ($i=1,\ldots,N_1$, $j=1,\ldots,N_2$, $k=1,\ldots,N_3$, say), $\rho^{PT}$ may be seen to possess three negative eigenvalue, namely, $-|c_1||c_2|\sqrt {\sum{p_i^2}\sum{q_j^2}}$, $-|c_2||c_3|\sqrt {\sum{q_j^2}\sum{r_k^2}}$, and $-|c_3||c_1|\sqrt {\sum{r_k^2}\sum{p_i^2}}$, where $c_1$, $c_2$, $c_3$ are the amplitudes of the three system states being measured. All these negative eigenvalues simultaneously go to zero in the environmental decoherence process we consider below. 
\section{Entanglement measure in terms of the nearest separable state}
\label{nss}
\noindent
As mentioned above, we call the kind of entanglement appearing in (1) `pure-mixed' entanglement wherein pure states of the system get correlated with mixed states of the apparatus; a likely consequence of environmental dephasing during the premeasurement interaction. This is to be distinguished from the more commonly considered pure-pure entanglement where only one of the $p$'s is unity, the others being zero, and similarly, only one of the $q$'s is unity. Thus a pure state 

\begin{equation}
|\phi\rangle =c_1|s_1\rangle\otimes|\phi_1^{(A)}\rangle+c_2|s_2\rangle\otimes|\phi_2^{(A)}\rangle
\label{SchPP}\end{equation}
would have given rise to the pure-pure density operator,

\begin{eqnarray}
	 \rho^{P-P}&=&|c_1|^2|s_1\rangle \langle s_1|\otimes |\phi_1^{(A)}\rangle\langle\phi_1^{(A)}|+
	|c_2|^2|s_2\rangle \langle s_2|\otimes|\phi_2^{(A)}\rangle\langle\phi_2^{(A)}|
\nonumber\\	&+&c_1c_2^*|s_1\rangle \langle s_2|\otimes |\phi_1^{(A)}\rangle\langle\phi_2^{(A)}|+c_1^*c_2|s_2\rangle \langle s_1|\otimes|\phi_2^{(A)}\rangle\langle\phi_1^{(A)}|,
\end{eqnarray}

\noindent where the superscript (P-P) is used to distinguish it from the pure-mixed case, and the similarity in the structures of $\rho$ and $\rho^{P-P}$ is apparent.
\vskip .5cm
\noindent
However, one important difference between the states $\rho$ and $\rho^{P-P}$ is that, for the latter, the Schmidt decomposition (Eq. \ref{SchPP}) is non-unique in the degenerate case $c_1=c_2$ \cite{Peres1993}, while no such non-uniqueness, in general, afflicts $\rho$. Thus, considering a transformation from $|s_1\rangle$, $|s_2\rangle$ to new states $|t_1\rangle$, $|t_2\rangle$, and correspondingly, from $|\phi_1^{(A)}\rangle$, $|\phi_2^{(A)}\rangle$ to, say, $|\psi_1^{(A)}\rangle$, $|\psi_2^{(A)}\rangle$ defined as
\vskip .5cm
\begin{eqnarray}
\pmatrix{|s_1\rangle\cr |s_2\rangle}=U\pmatrix{|t_1\rangle\cr |t_2\rangle},~~\pmatrix{|\phi_1^{(A)}\rangle\cr |\phi_2^{(A)}\rangle}=U^*\pmatrix{|\psi_1^{(A)}\rangle\cr |\psi_2^{(A)}\rangle}
\end{eqnarray}   
\vskip .5cm 
\noindent 
where $U$ is a $2$x$2$ unitary matrix, one finds that (Eq. \ref{SchPP}) continues to hold with $|s_i\rangle$, $|\phi_i^{(A)}\rangle$ replaced with $|t_i\rangle$, $|\psi_i^{(A)}\rangle$ ($i=1,2$; higher dimensional generalization is straightforward). Evidently, such a unitary transformation is ruled out for $\rho$ for unequal dimensions $N_1$, $N_2$ relating to the two  groups of microstates of A, which is expected to be the generic situation in the measurement context.

\vskip .5cm
\noindent
As we see below, one result of the environmental perturbations on the apparatus A is to bring about a maximal mixing among the individual groups of apparatus microstates and so, it is really not relevant as to what the sets of weights $p_i$ and $q_j$ are to start with. Additionally, the process of environment-induced decoherence leads to an evolution of the S-A density matrix towards the {\it nearest} separable state, wherein the off-diagonal blocks in $\rho$ get erased, leaving a classically correlated S-A state.  The fact that no other representation of the form (Eq. 1) is possible for the S-A entangled state is possible, coupled with this tendency of decoherence towards the nearest separable state leads one to apartial resolution of the so-called preferred basis problem while there remains the problem relating to the stability of the pointer states (see section IV below). 
\vskip .5cm
\noindent
It has been shown \cite{Vedral1997,Vedral1998} that a measure of quantum entanglement of a composite state is obtained by referring to its distance (in terms of an appropriate distance function) from the nearest separable state, a convenient distance function between two density operators being the relative entropy defined below.
In the case of $\rho^{P-P}$ the nearest separable state turns out to be $\sigma^{P-P}$ \cite{Vedral1998} given by
 
\begin{equation}
\sigma^{P-P}={\mathrm{diag}}(\rho^{P-P}),
\end{equation}
\noindent
where `$\mathrm{diag}$' stands for the matrix containing only the diagonal blocks of the matrix representing $\rho^{(P-P)}$ in the Schmidt basis. Although the relative entropy is neither symmetric nor satisfies the triangle inequality, it has properties (see e.g.~\cite{Nielsen2002}) that make it useful as a distance function. We will demonstrate in a manner essentially similar to ~\cite{Vedral1998} that $\rho^*\equiv{\mathrm{diag}}(\rho$) is the nearest disentangled state to $\rho$ in the pure-mixed case as well. The relative entropy is defined as,
\begin{equation}
s(\rho|\rho^*)={\mathrm{Tr}}\rho[\ln\rho-\ln\rho^*],
\label{relent}\end{equation}
\noindent
and, referring to
\begin{equation}
f(x)=s(\rho|(1-x)\rho^*+x\sigma),
\end{equation}
\noindent as the relative entropy of $\rho$ and a convex combination of $\rho^*$ with an arbitrary separable state $\sigma$, the proof depends on showing that $df(x)/dx|_{x=0}\geq 0$. The analysis is local in the sense that among all the separable states in a neighbourhood of $\rho^*$, the one nearest to $\rho$ is $\rho^*$ itself.
\vskip .5cm
\noindent
Using the integral representation,
\begin{equation}
\ln a=\int_0^{\infty}{at-1\over a+t}{dt\over 1+t^2},
\end{equation}
we can write,
\begin{eqnarray}
\left.{df(x)\over dx}\right|_{x=0}&=&\lim_{x\rightarrow 0}[\frac{f(x)-f(0)}{x}]\nonumber\\& = &1- \int_0^{\infty}{\mathop{Tr}[(\rho^*+t)^{-1}\rho
(\rho^*+t)^{-1}\sigma]dt}.
\end{eqnarray}
\noindent Recognizing that $\rho^*$ is diagonal in the basis under consideration, and calling its eigenvalues $\lambda_j$, the integrand can be written as,
\begin{displaymath}
\sum_{j,k}{1\over (\lambda_j+t)(\lambda_k+t)}\rho_{j,k}\sigma_{k,j},
\end{displaymath}
\noindent by making use of the resolution of identity in terms of the eigenvectors of $\rho^*$.
\vskip .5cm
\noindent
Interchangeing the order of integration and summation, integration of the diagonal terms  yields
\begin{displaymath}
\sum_{j=1}^{N_1}\sigma_{j,j}+\sum_{k=2N_1+N_2+1}^{2N1+2N_2}\sigma_{k,k},
\end{displaymath}
\noindent while for the off-diagonal terms we get the contribution
\begin{displaymath}
\sum_{j\neq k}{1\over \sqrt {\lambda_j\lambda_k}}g(\lambda_j,\lambda_k)\rho_{j,k}\sigma_{k,j},
\end{displaymath}
where, the function $g(\lambda_j,\lambda_k)$, defined for  $\lambda_j,~\lambda_k \in [0,1]$ as,
\begin{displaymath}
g(\lambda_j,\lambda_k)={\sqrt{\lambda_j\lambda_k}\over \lambda_j-\lambda_k}\ln {\lambda_j\over\lambda_k},
\end{displaymath}
\noindent is limited by $0\leq g(\lambda_j,\lambda_k)\leq 1$. Before carrying out the summations involved in the trace, we assemble all the elements we need. The non-zero off-diagonal matrix elements of $\rho$ are,
\begin{eqnarray}
\rho_{j,k}&=&c_1c_2^*p_j q_k,\hspace{.5cm}\mbox{for}\hspace{.2cm} j=1,...,N_1;\hspace{.2cm}k=2N_1+N_2+1,...,2N_1+2N_2,\nonumber\\&=&c_1^*c_2q_j p_k,\hspace{.5cm}\mbox{for}\hspace{.2cm} j=2N_1+N2+1,...,2N_1+2N_2;\hspace{.2cm}k=1,...,N_1,
\end{eqnarray}
\noindent and the eigenvalues of $\rho^*$ are given by,
\begin{eqnarray}
\lambda_j&=&|c_1|^2p_j,\hspace{.5cm}\mbox{for}\hspace{.2cm} j=1,...,N_1,\nonumber\\&=&|c_2|^2q_j,\hspace{.5cm}\mbox{for}\hspace{.2cm}j=2N_1+N_2+1,...,2N_1+2N_2.
\end{eqnarray}
\vskip .5cm
\noindent The disentangled state $\sigma$ is taken in a product form viz.
\begin{eqnarray}
\sigma=(|\alpha\rangle\langle\alpha|)\otimes(|\beta\rangle\langle\beta|),
\end{eqnarray}
where the first and second terms belonging to the system and apparatus respectively are given by, say, 
\begin{eqnarray}
|\alpha\rangle=\sum_{i=1}^2d_i|s_i\rangle,\hspace{.5cm}|\beta\rangle=\sum_{i=1}^{N_1+N_2}f_i|u_i\rangle,
\end{eqnarray}
\noindent with $|u_i\rangle$ being $|a_i\rangle$ for $i$ in the range $1$ to $N_1$ and  $|b_i\rangle$ otherwise (more general separable states in the form of sum over products need not be considered separately, see~\cite{Vedral1998}). The matrix for $\sigma$ now has a block form with the diagonal blocks being given by $|d_1|^2\sum_{i,j}f_if_j^*$ and $|d_2|^2\sum_{i,j}f_if_j^*$ and the off diagonal blocks by $d_1^*d_2\sum_{i,j}f_if_j^*$ and its complex conjugate respectively with $i$ and $j$ running over $1$ to $N_1+N_2$.
\vskip .5cm
\noindent
Putting all this together, we now evaluate the sum involved in the calculation of the trace with the result,
\begin{eqnarray}
 {df(x)\over dx}|_{x=0}&=&1-[|d_1|^2\sum_{i=1}^{N_1}|f_i|^2+|d_2|^2\sum_{i=N_1+1}^{N_1+N_2}|f_i|^2 \nonumber\\&+&\sum_{j=1}^{N_1}\sum_{k=1}^{N_2}{c_1c_2^*\sqrt{p_jq_k}\over |c_1||c_2|}g\{|c_1|^2p_j,|c_2|^2q_k\}d_1d_2^*f_{N_1+k}f_j^*+c.c.],
\end{eqnarray}
\noindent where $c.c.$ stands for the complex conjugate of the third term in the brackets. Taking absolute values and remembering that the function $g$ lies in the range $0$ to $1$ we can show that the modulus of the square bracket is limited by $|d_1|^2+|d_2|^2$ and hence by $1$ which means that the relevant derivative is positive. We thereby conclude that even for the pure-mixed state $\rho$ considered above, the nearest disentangled state is $\rho^*$ i.e. the state obtained from $\rho$ by the removal of the off diagonal blocks.
\vskip .5cm
\noindent
For pure states of bipartite systems the relative entropy of entanglement   $s(\rho|\rho^*)$ reduces to the von Neumann entropy which is the usual measure of entanglement. If we define $\rho_S$ and $\rho_A$ to be the reduced density matrices for the system and apparatus respectively then, generally speaking, the total correlation ( or mutual information) is given by $s(\rho|\rho_S^*\otimes \rho_A^*)$ of which $s(\rho|\rho^*)$ is the quantum part and $s(\rho^*|\rho_S^*\otimes \rho_A^*)$ is the classical part \cite{Henderson2001} although we do not imply any additivity of the parts in making the whole. 
\vskip .5cm
\noindent
To get an idea of the magnitudes of such correlations let us consider a solvable case where $p_i=1/N_1$ and $q_j=1/N_2$ for all values of the indices, i.e., let us assume $\rho_a^{(A)}$ and $\rho_b^{(A)}$ to be maximally mixed states in the respective subspaces. The non-zero eigenvalues of $\rho$ are then $\frac{|c_1|^2}{N_1}$ ($(N_1-1)$-fold degenerate), $\frac{|c_2|^2}{N_2}$ ($(N_2-1)$-fold degenerate) and $\frac{N_1-|c_1|^2(N_1-N_2)}{N_1N_2}$, and therefore,

\begin{equation}
s(\rho|\rho^*)={|c_1|^2\over N_1}\ln[1+{|c_2|^2N_1\over |c_1|^2N_2}]+{|c_2|^2\over N_2}\ln[1+{|c_1|^2N_2\over |c_2|^2N_1}],
\end{equation}
\begin{equation}
s(\rho^*|\rho_S^*\otimes \rho_A^*)=-|c_1|^2\ln |c_1|^2-|c_2|^2\ln |c_2|^2.
\label{clcor}\end{equation}
\vskip .5cm
\noindent
Thus, the classical part of the correlation in the pure-mixed case is just the von Neumann entropy of one of the subsystems while the quantum part is suppressed by the reciprocals of the degeneracy factors. One observes that the quantum correlation decreases towards zero for large values of $N_1$, $N_2$. For a {\it pure} initial state $\rho$, on the other hand, the quantum and classical correlations are {\it both} given by (\ref{clcor}); (see~\cite{Zyckowski2001} for estimates for bipartite entanglement of arbitrarily chosen pure states) which again points to the crucial role played by mixing in the relative measures of quantum and classical correlations.
\vskip .5cm
\noindent
In other words, the macroscopic nature of the pointer states is seen to imply a drastic reduction in the degree of quantum entanglement as a result of which, this part of the total correlation gets removed in the environment-induced decoherence in quite a short time (see, e.g.,~\cite{Allahverdyan2007} for an estimate of the decoherence time for a macroscopic measuring apparatus). 
\vskip .5cm
\noindent 
In the following we shall need, in addition to the nearest separable state $\rho^{*}$, the state $\rho_0$ obtained from $\rho^{*}$ by assigning identical values ($\frac{1}{N_1}$)to all the $p$'s and also identical values ($\frac{1}{N_2}$)to all the $q$'s. Evidently, $\rho_0$ involves a classical correlation between the system states $|s_1\rangle$, $|s_2\rangle$ and the equimixed apparatus states  

\begin{equation}
\sigma_a^{(A)}\equiv{1\over N_1}{\sum_{i=1}^{N_1}|a_i\rangle\langle a_i|}
\end{equation}
and
\begin{equation}
\sigma_b^{(A)}\equiv{1\over N_2}{ \sum_{i=1}^{N_2}|b_i\rangle\langle b_i|}.
\end{equation}
\vskip .5cm 
\noindent 
The classical correlation is characterized by the absence of off-diagonal terms in $\rho_0$:

\begin{equation}
\rho_0= |c_1|^2|s_1\rangle \langle s_1|\otimes \sigma_a^{(A)}+
|c_2|^2|s_2\rangle \langle s_2|\otimes \sigma_b^{(A)}.
\end{equation}.

\section{The decoherence process}
\label{dcp}
\noindent We now turn to the description of the decoherence mechanism. The total Hamiltonian we consider is given by ( for the sake of generality, the dimensions of the state spaces of the three systems are denoted by $N_S, N_A$ and $N_E$ respectively; as indicated above, we use $N_S=2, N_A= N_1+N_2$, with appropriate choices for $N_1, N_2$, as also for $N_E$, see below),
\begin{equation}
H=H_S\otimes I_A\otimes I_E+I_S\otimes H_A\otimes I_E+I_S\otimes I_A\otimes H_E+\lambda I_S\otimes V_{A-E}.
\label{totalH}\end{equation}
Here $I_S,~ I_A,~ I_E$ denote identity operators for S, A and E respectively,  $H_S,~H_A,~ H_E$ represent the Hamiltonians for S, A and E considered in isolation,  the latter two being diagonal in the respective sets of basis states chosen, and $V_{A-E}$ stands for the A-E interaction responsible for the decoherence, with strength $\lambda$, which we assume to be small (weak coupling limit) in the present context. A few relevant aspects of the Hamiltonian (\ref{totalH}) are discussed in \cite{Lahiri2007}, and in this context we make the following observations.
\vskip .5cm
\noindent
(i) The crucial assumption underlying our results is that the pointer states of the apparatus have well-defined energy values. We note that it is the pre-measurement interaction that selects out, among all possible dynamical variables of the apparatus, a particular one that constitutes the pointer variable for the measurement under consideration. It is, in principle, possible to have a pre-measurement interaction Hamiltonian effecting this selection regardless of the specific features of the apparatus. However, only a specific class of macroscopic systems can qualify as the measuring apparatus for a given measurement. We assume that one additional requirement to be satisfied by the apparatus is that the pointer variable selected out by the pre-measurement interaction has to commute with the apparatus Hamiltonian. It is worthwhile to explore the conjecture that there exists a class of measurements where this has to be a necessary feature of the pre-measurement interaction and of the system A if the latter is to constitute an appropriate apparatus effecting the measurement under consideration. Referring to von Neumann's measurement scheme, for instance, where one measures the spin of S by means of the momentum of a free particle, which constitutes the pointer variable \cite{Peres1993}, or, equivalently, the Stern-Gerlach measurement scheme considered in \cite{Venugopalan1995}, one observes that the pre-measurement interaction involving the position co-ordinate of the apparatus selects out the apparatus momentum as the pointer variable, which does commute with the apparatus Hamiltonian. A similar situation obtains in the measurement of a spin with an Ising magnetic dot \cite{Allahverdyan2007}. On the other hand, there may exist measurement set-ups (for instance, one in which the apparatus for measuring the spin is a harmonic oscillator \cite{Venugopalan1999}, with the position variable of the oscillator being involved in the S-A interaction) where the pointer variable does not necessarily commute with the apparatus Hamiltonian but is still stable against environmental perturbations. While instances of the latter type are not covered by our work, it is nevertheless possible that the macroscopic nature of the apparatus \cite{Peres1993,Frasca2006} ensures an effective energy conservation for pointer states where the weak environmental perturbations fail to cause transitions between the latter.

\vskip .5cm
\noindent
(ii) A relevant question relates to the choice for the A-E interaction operator $V_{A-E}$. Since the measuring device is a macroscopic system, the most realistic choice should be an operator represented by a random Hermitian matrix in an arbitrarily chosen basis; indeed, any other form would imply some special assumption or other relating to the interaction and would be contrary to the macroscopic nature of the measuring device and the environment. These  are effectively classical systems \cite{Peres1993} with densely bunched degenerate states whose interactions are, generically speaking, chaotic in nature. The quantum features of such interactions are known to be similar to those of ensembles of random matrices. A large body of recent work has looked into the entangling power of chaotic interactions (see, e.g.~\cite{Lakshminarayan2001,Scott2004,Ghosh2004}), and a number of these also bring out random features in the density matrix fluctuations in subsystems interacting with one another through such random matrices \cite{Lahiri2007,Nag2005} where one finds that the chaotic interactions are effective in reducing the states of the subsystems to classical mixtures. 
\vskip .5cm
\noindent
As seen from the numerical evidence below, the Hamiltonian (\ref{totalH}) with randomly chosen matrix elements does efficaciously disentangle the S-states from the A-states, leading to a state in which only the classical correlations between the two remain; for such a state one can talk in terms of `pre-existing' properties in S. One can describe the decoherence process as one of entanglement sharing (see, e.g., \cite{Koffman2000}) in the tripartite S-A-E system (\cite{Vedral2003} presents a bound relating the information gained in the measurement and the over-all mixedness of the apparatus state; while we consider equimixing among two distinct groups of the apparatus states, maximal mixing among {\it all} the apparatus microstates would render it incapable of effecting the measurement). Environment-induced decoherence (see~\cite{Schlosshauer2004} for a review) in bipartite composite systems where the environment acts directly only on one of the two subsystems, has been considered in \cite{Lombardo2005}. 
\vskip .5cm
\noindent
The reduced S-A density matrix ($\rho^{{S-A}}$) elements fluctuate during the process, whereby $\rho^{S-A}$ undergoes a Brownian-like motion in the space of entangled states, tending to the nearest separable state $\rho^*$, while at the same time deviating from the latter due to  mixing among the two groups of apparatus states alluded to above, finally reaching the state $\rho_0$.

\vskip .5cm
\noindent
A pure decoherence process (i.e. one without mixing among the relevant groups of apparatus microstates) can however be generated by a non-demolition type coupling between the apparatus and the environment wherein the interaction term is taken to be a product of a function of the apparatus Hamiltonian and a random matrix in the environmental space. The evolution does not affect the diagonal terms of the reduced density matrix while its off diagonal terms get erased; the final state being described by $\rho^*$ rather than $\rho_0$. 
 
\section{Numerical Evidences}
\label{nume}
\noindent We introduce the following measures for decoherence and for the equimixing among groups of apparatus microstates indicated above (where the latter may be looked upon as the early stage of the relaxation process in the state space of the apparatus) :
\begin{equation}
Q_D(t) \equiv \sum_{i\leq N_1 \atop j> 2N_1+N_2}|\rho_{i,j}^{S-A}(t)|^2,
\end{equation}

\noindent i.e., the sum of the modulus squared of the off diagonal elements of the reduced system-apparatus density matrix at time $t$, and
\begin{equation}
Q_R(t) \equiv \sum_{i=1}^{N_1}\left|{1\over N_1}-\rho_{ii}^{S-A}(t)\right|^2+\sum_{i=2N_1+N_2+1}^{2(N_1+N_2)}\left|{1\over N_2}-\rho_{ii}^{S-A}(t)\right|^2,
\end{equation}

\noindent i.e. the sum of the squared deviations of the diagonal elements from their respective equimixed values. According to the picture outlined above, these quantities are expected to approach zero with time, apart from fluctuations caused by the finiteness of the dimensions involved in the computation.
  
\begin{figure}[htb]
	\centering
		\subfigure[ Random matrix coupling ]{\label{fig:decoh1}\includegraphics[width=0.4\textwidth]{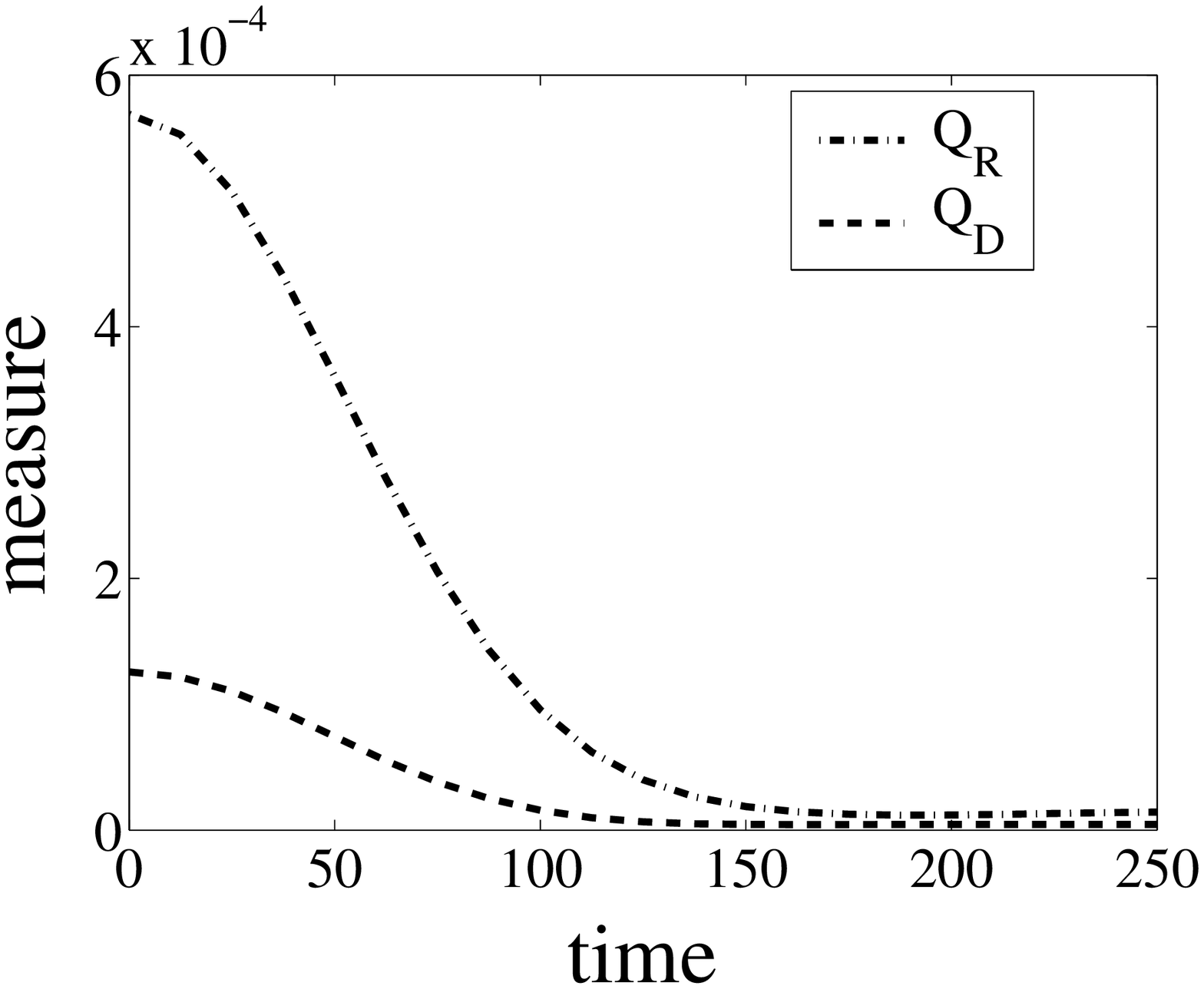}}\qquad
  \subfigure[Nondemolition coupling]{\label{fig:decoh2}\includegraphics[width=0.4\textwidth]{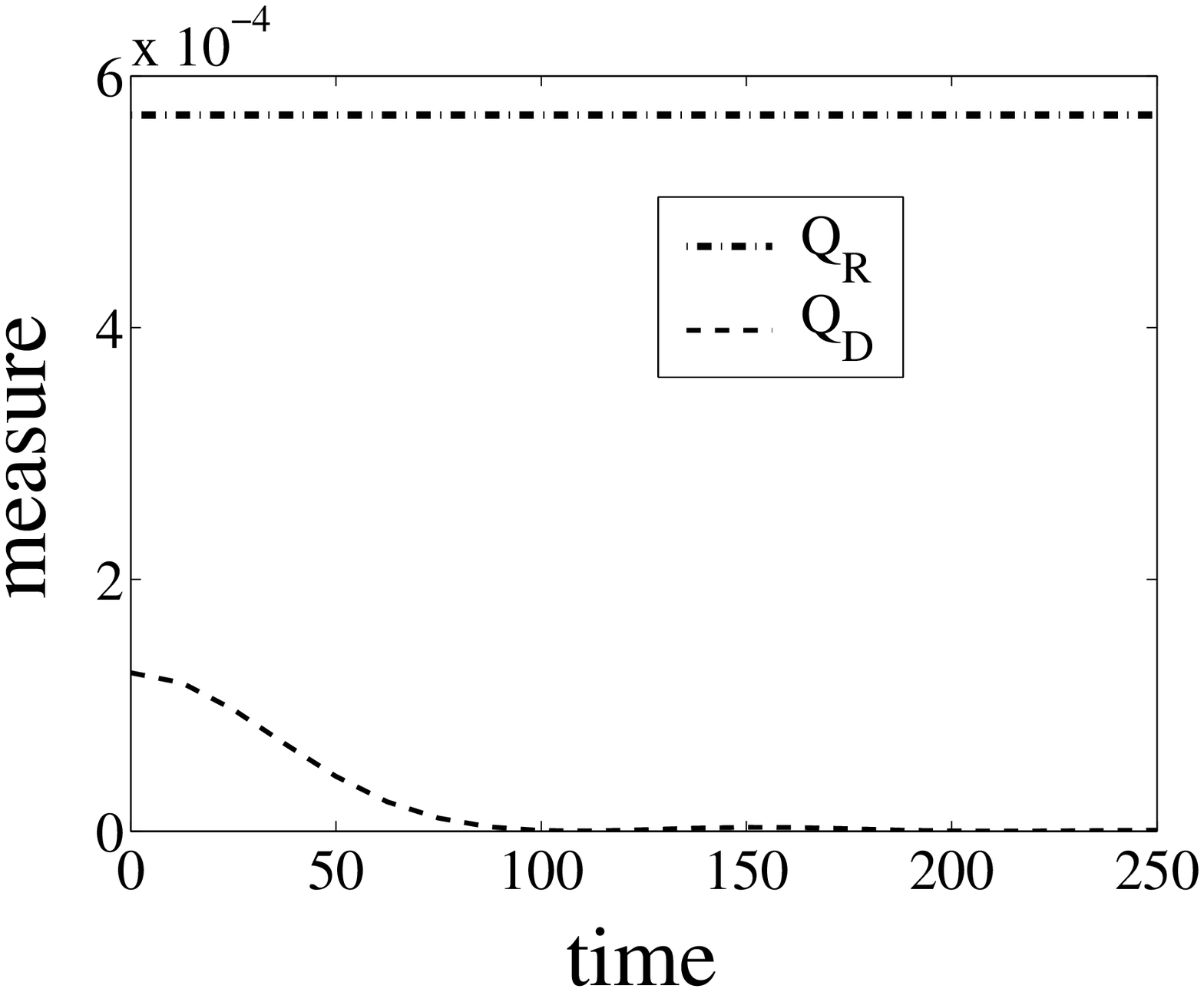}}
\caption{Decoherence and relaxation with (a) $V_{A-E}=V_R^{A-E}$, and (b)$V_{A-E}=H_A\otimes V_R^E$ (see text) with $N_S=2$,    $N_1=7$, $ N_2=8$, $ N_E=60$; the energy eigenvalues of A (with degeneracies $N_1,~N_2$) are $E_a=200.0,\;E_b=400.0$; $H_S$ is a $2\times 2$ matrix with all four elements set at $0.5\times 10^{-6}$, and so does not commute with $\hat S$; $\;c_1=1/\sqrt{2};\;
c_2=1/\sqrt{2}$; $\lambda=0.005$ in (a) and $\lambda=0.0001$ in (b); $H_E$ is chosen diagonal with eigenvalues spread uniformly in the range $190-410 $; for notations see text.}
\end{figure}
\begin{figure}[htb]
	\centering
		\includegraphics[width=0.4\textwidth]{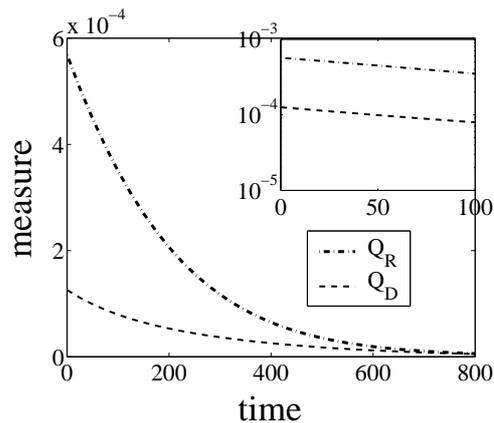}
		\caption{Decoherence and relaxation with infinite bath simulation; $V_{A-E}=V_R^{A-E}$; parameter values are the same as in figures~\ref{fig:decoh1}; the exponential fall is apparent from the semi-logarithmic plot in the inset.}
	\label{fig:decoh3}
\end{figure}

\vskip .5cm
\noindent
Figures~\ref{fig:decoh1} and \ref{fig:decoh2} show the time variation of these two indices for (a) an A-E coupling represented by a real Hermitian matrix $V_{A-E}=V_R^{A-E}$ with randomly selected elements and (b) a non-demolition type coupling of the form $V_{A-E}= H_A\otimes V_R^E$ where $V_R^E$ stands for an operator in the state space of E represented by a real Hermitian matrix once again with randomly chosen elements. One observes essentially similar features in decoherence in the two situations, although relaxation is absent in the latter. The Gaussian feature of the decoherence process apparent from the figures, and verified quantitatively from the numerical data, is an artifact arising from the finiteness of the environment and the resulting memory effects in the decoherence process. One can eliminate the memory effect by renewing the environment state after each time step, thus simulating an unchanging and therefore infinite bath (see~\cite{W.H.Zurek2003} for possible time-courses of decoherence). As seen from Fig.~\ref{fig:decoh3}, this results in an exponential rather than Gaussian decay for the above mentioned measures. 
\vskip .5cm
\noindent 

\begin{figure}[htb]
	\centering
		\includegraphics[width=0.45\textwidth]{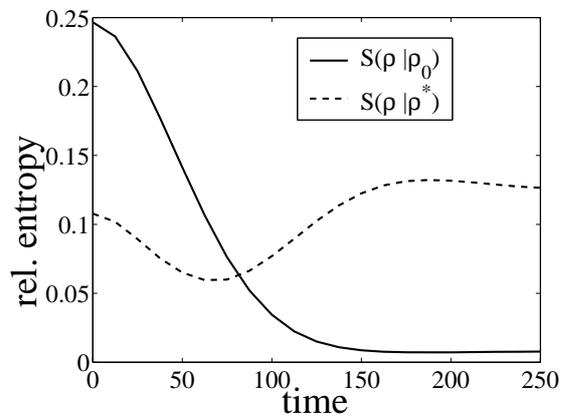}
	\caption{The relative entropy between $\rho^{S-A}(t)$ and $\rho^*$, and between  $\rho^{S-A}(t)$ and $\rho_0$; $V_{A-E}=V_R^{A-E}$; other parameters same as in Fig.~\ref{fig:decoh1}}.
	\label{fig:RelEn}
\end{figure}

\vskip .5cm
\noindent Figure~\ref{fig:RelEn} depicts the time-variation of the relative entropy, defined in (Eq.~\ref{relent}), between $\rho^{S-A}(t)$ and $\rho^*$ as also between $\rho^{S-A}$ and $\rho_0$, and clearly shows that $\rho^{S-A}$ first approaches $\rho^*$ due to the process of environment-induced decoherence, but then deviates from the latter, veering instead towards $\rho_0$ due to equimixing among groups of pointer states as part of the relaxation process.

\vskip .5cm
\noindent Analogous to the relative entropy, the Bures metric \cite{Vedral1998} provides one with another distance function and is defined by,
\begin{eqnarray}
D_B(\rho|\rho^*)=2-2\sqrt {F(\rho|\rho^*)}
\end{eqnarray}

\noindent where $F(\rho|\rho^*)=[{\mathop{Tr}}(\sqrt{(\rho^*)}\rho\sqrt{(\rho^*)})^{1/2}]^2$ denotes the fidelity distance between $\rho$ and $\rho^*$. 
\vskip .5cm

\begin{figure}[htb]
	\centering
		\subfigure[Random matrix coupling]{\label{fig:fidelity}\includegraphics[width=0.40\textwidth]{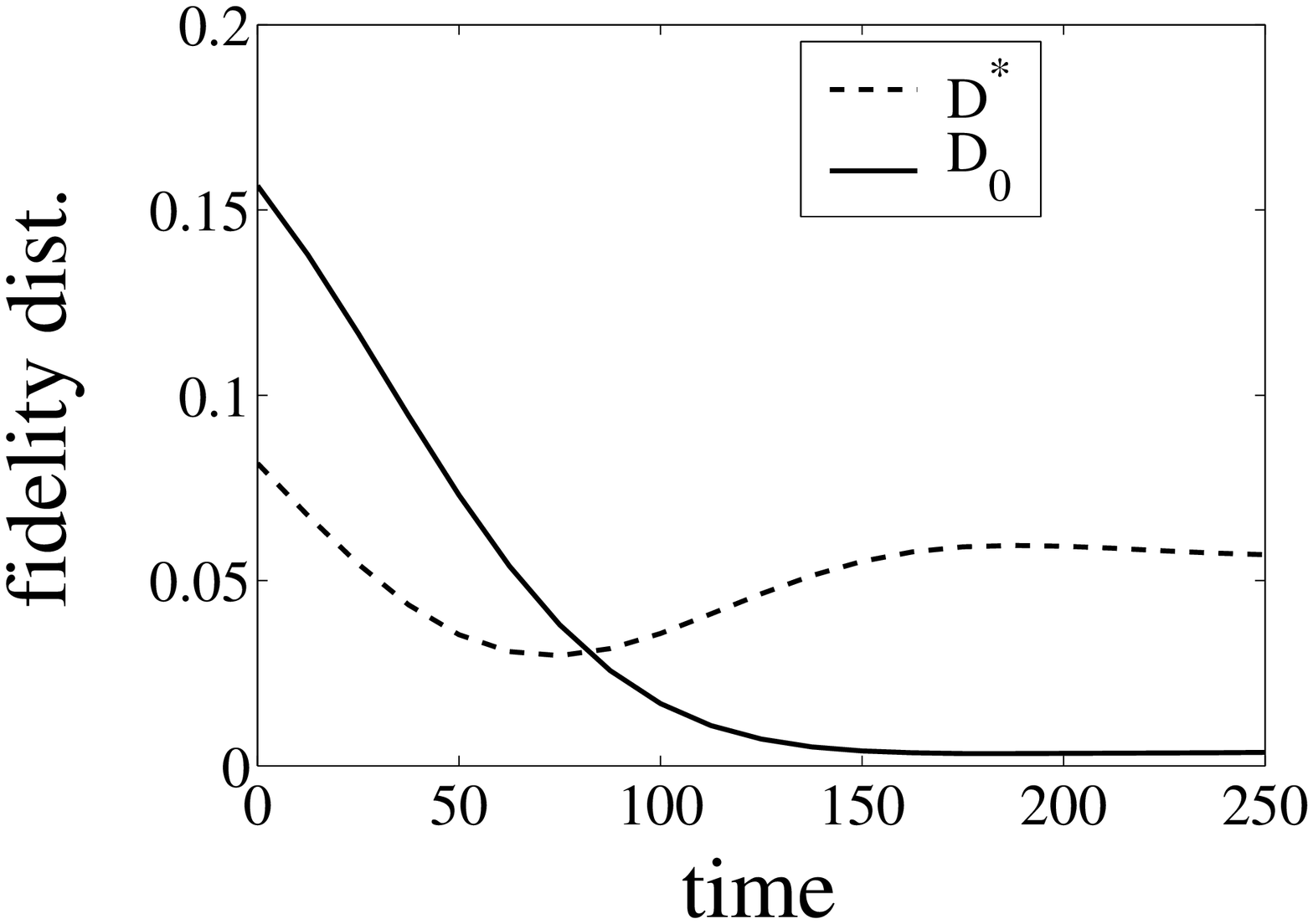}}\hskip 2mm
		\subfigure[Nondemolition coupling]{\label{fig:fidelitydis}\includegraphics[width=0.40\textwidth]{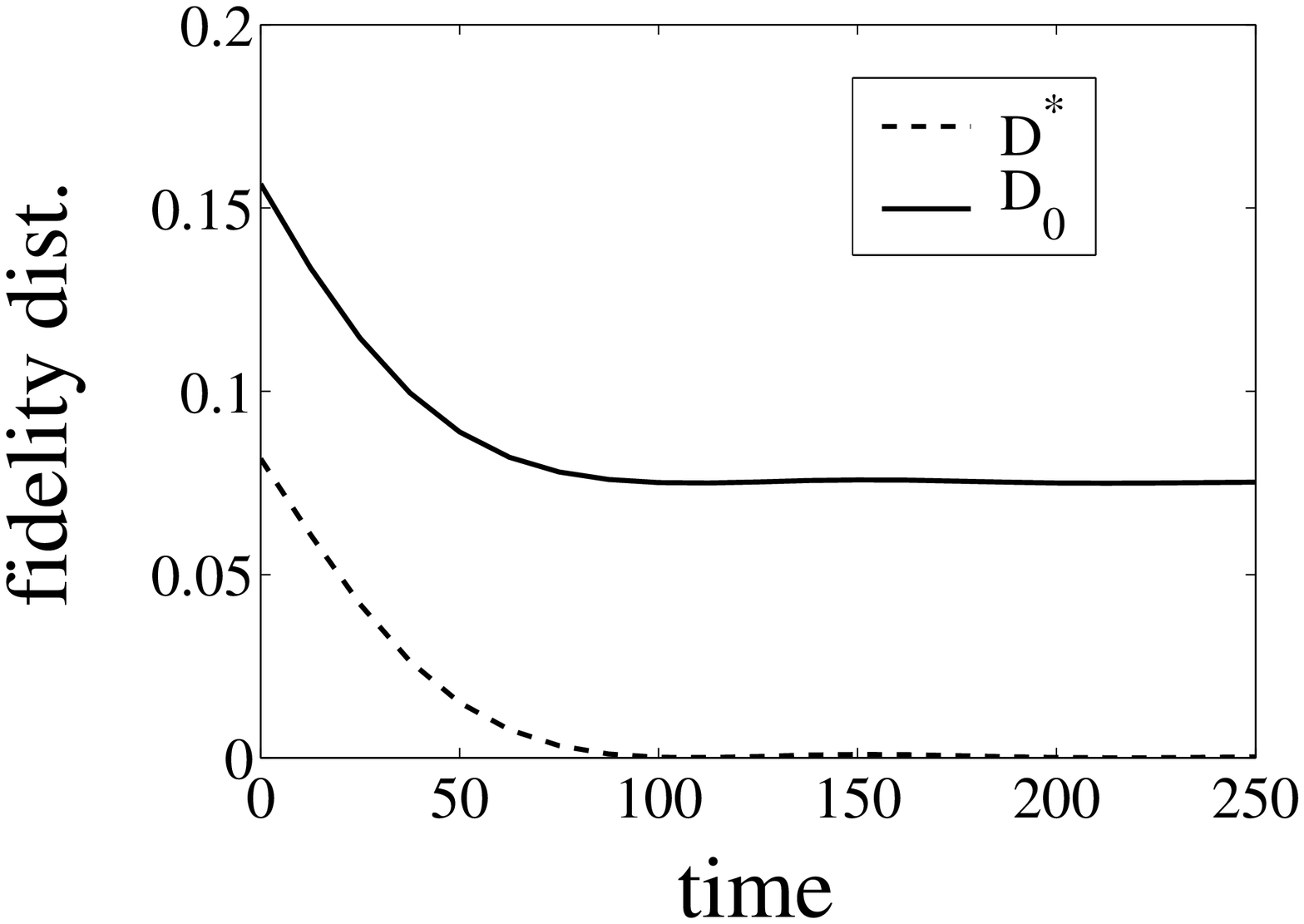}}
	\caption{Variations of Bures distances $D_0$ and $D^*$ (see text) with time for (a) $V_{A-E}=V_R^{A-E}$, and (b) $V_{A-E}=H_A\otimes V_R^E$; parameters same as in Figures~\ref{fig:decoh1} and \ref{fig:decoh2} .}	
\end{figure}

\noindent  The variations of Bures distance between $\rho^{S-A}(t)$ and $\rho_{0}$ ($D_0$), and that between $\rho^{S-A}(t)$ and $\rho^{*}$ ($D^*$) with time, shown in (Fig.~\ref{fig:fidelity}) for $V_{A-E}=V_R^{A-E}$,  clearly depict the decoherence and mixing processes as outlined above. In the long run, $\rho^{S-A}(t)$ tends to coincide with $\rho_{0}$, and maintains a steady separation from the nearest disentangled state $\rho^{*}$.

\vskip .5cm
\noindent  For a non-demolition coupling, $V_{A-E}=V_R^{A-E}$, on the other hand, the distance  from $\rho^{*}$ ($D^*$) diminishes to zero while $D_0$ tends to a steady non-zero value since the diagonal terms remain unaltered (Fig.\ref{fig:fidelitydis}). This confirms that in the absence of the relaxation process $\rho^{S-A}$ does indeed tend to the nearest disentangled state due to decoherence alone. 

\begin{figure}[htb]
	\centering
		\includegraphics[width=0.4\textwidth]{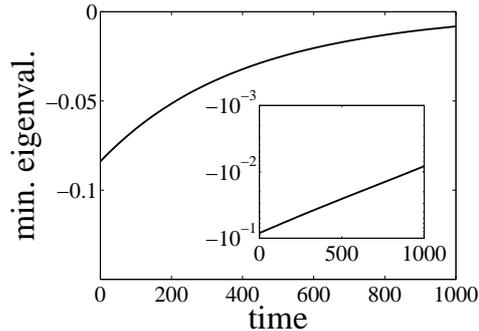}
		\caption{Variation of the minimum eigenvalue of the partial transpose of $\rho^{S-A}$ with time; parameters same as in Fig.~\ref{fig:decoh1}; (inset) semi-log plot.}
	\label{fig:eig}
\end{figure}

\vskip .5cm
\noindent Finally, Figure~\ref{fig:eig} depicts the time variation of the minimum eigenvalue of the partial transpose of $\rho^{S-A}(t)$. As already explained in section II, the partial transpose of the initial S-A state ($\rho$) possesses exactly one negative eigenvalue. Our numerical results show that $\rho^{S-A}(t)$ also possesses a single negative eigenvalue during the entire course of decoherence (this has non-trivial implications regarding the decoherence process we shall indicate elsewhere) which tends to zero with time. As the quantum correlations between S and A are erased and the S-A state becomes separable, the minimum eigenvalue continues to remain zero.

\section{Outlook and concluding remarks}
\label{ocr}
\noindent While our presentation evokes a measurement context, the main results are amenable to an independent appraisal. Indeed, the measurement model used here is based on a number of simplifying assumptions and is at best of limited applicability. For instance, the pre-measurement interaction and the decoherence process resulting from environmental dephasing are not temporally distinct and independent processes and actually proceed simultaneously (see, e.g.~\cite{Venugopalan1999,Venugopalan1995}). In the present paper we view these as {\it logically} distinct processes and have looked at the consequence of the latter process on the outcome of the former when these are {\it temporally} distinct as well. Moreover, we do not explain why the pointer states pertaining to the measurement under consideration are stable under environmental perturbations, assuming instead that they have to correspond to well-defined energy values of the apparatus if the measurement is to succeed. While this may well be so because of specific features of the measuring apparatus (a specific measurement requires not just any apparatus but a specific one) including the fact that the latter is a macroscopic system, we have not looked into details of the underlying mechanism in the present paper.
\vskip .5cm
\noindent
On the other hand, independently of the measurement context, our work extends results in \cite{Vedral1998} relating to entanglement measure of states of composite systems involving quantum correlations. Further, it provides a comparison of classical and quantum correlations for a class of states of a microscopic system S correlated with a macroscopic system A, and indicates that such a state approaches the nearest separable state under decoherence. Possible implications of these results for the measurement problem constitute a separate consideration in this paper.
\bibliography{decoh}

\end{document}